\documentclass[11pt,showpacs,preprintnumbers,amsmath,amssymb,prd,nofootinbib,superscriptaddress]{revtex4-2}

\usepackage{dcolumn}
\usepackage{bm}
\usepackage{ifpdf}
\usepackage{hyperref}
\usepackage{bm}
\usepackage{xcolor,color,graphicx,graphics,physics}
\usepackage[spanish,english]{babel}
\usepackage[latin1]{inputenc}
\usepackage[OT1]{fontenc}
\usepackage{latexsym,amssymb,amsmath,amsfonts, slashed}
\usepackage{makeidx}
\usepackage{epsfig,subfigure}
\usepackage{natbib}
\usepackage{epstopdf}
\usepackage{mathrsfs}
\usepackage{hyperref}
\hypersetup{colorlinks=true, linkcolor=blue, citecolor=blue, urlcolor=blue}
\usepackage{enumerate}

\usepackage{fixmath}

\everymath{\displaystyle}
\usepackage{graphicx}

\usepackage[T1]{fontenc}
\usepackage{amsmath}
\usepackage{amssymb}
\usepackage{graphicx}
\usepackage{xcolor}

\newcommand{\bea}{\begin{eqnarray}}
\newcommand{\eea}{\end{eqnarray}}

\newcommand{\orcid}[1]{\href{https://orcid.org/#1}{\includegraphics[width=10pt]{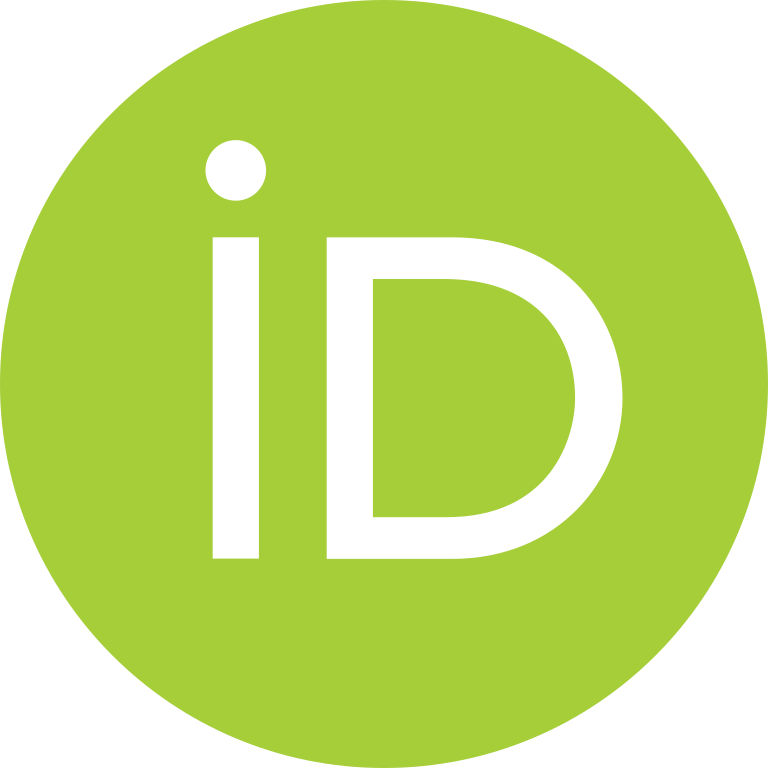}}}

\begin{document}

\title{Violation of Lorentz symmetries and thermal effects in Compton scattering}

\author{D. S. Cabral  \orcid{0000-0002-7086-5582}}
\email{danielcabral@fisica.ufmt.br}
\affiliation{Instituto de F\'{\i}sica, Universidade Federal de Mato Grosso,\\
78060-900, Cuiab\'{a}, Mato Grosso, Brazil}

\author{A. F. Santos \orcid{0000-0002-2505-5273}}
\email{alesandroferreira@fisica.ufmt.br}
\affiliation{Instituto de F\'{\i}sica, Universidade Federal de Mato Grosso,\\
78060-900, Cuiab\'{a}, Mato Grosso, Brazil}

\author{Faqir C. Khanna \orcid{0000-0003-3917-7578} \footnote{Professor Emeritus - Physics Department, Theoretical Physics Institute, University of Alberta\\
Edmonton, Alberta, Canada}}
\email{fkhanna@ualberta.ca; khannaf@uvic.ca}
\affiliation{Department of Physics and Astronomy, University of Victoria,\\
3800 Finnerty Road, Victoria BC V8P 5C2, Canada}

\begin{abstract}

In this paper, the differential cross section for the Compton scattering process is calculated. Two types of corrections are investigated: corrections due to violation of Lorentz symmetry and thermal effects. An extended QED is considered to introduce the parameter that leads to the breaking of symmetry. While temperature effects are introduced using Thermofield Dynamics formalism. It is shown that the differential cross section changes with both corrections. These corrections are dominant at appropriate limits. These special cases are analyzed and compared with other results from the literature.

\end{abstract}

\maketitle

\section{Introduction}

Quantum Electrodynamics (QED) is the theory that deals with the interaction between elementary particles, such as fermions and photons. Compton scattering \cite{comptonwoo} is an example of the many scattering processes explained by QED. This process consists of an interaction between electrons and photons through an intermediate electron.  The QED is a sector of the Standard Model (SM), which is considered the most complete theory about nature. The SM is built based on Lie group theories, having as its primary structure the product $SU(3)\times SU(2)\times U(1)$ \cite{gaillardmodelopadrao}. The SM describes elementary particles, their fundamental properties and their interactions, except gravity \cite{kane,moreiramodelo,novaes}. The fact that SM does not include gravity makes this theory incomplete with respect to the description of matter and its interactions. Such a situation suggests that SM is not fundamental, but effective. However, there are some theories that propose to deal with all fundamental forces, such as the string theory and supersymmetric models \cite{moreiramodelo, Mukhi}, among others.

In string theory, Kosteleck\'y and Samuel studied a process similar to the Higgs mechanism, looking for components of the tensor field that have a non-zero vacuum expectation value. It was then seen that this fact leads to a spontaneous violation of Lorentz symmetry in the theory \cite{kosteleckycordas}. A year later, Carroll, Field and Jackiw studied the breaking of Lorentz and parity symmetries introducing a background vector into the theory. They used astrophysical data to calculate possible observational limits of the parameters that characterize such a violation \cite{carrollfieldjackiw}. From that, Colladay and  Kosteleck\'y introduce an extension to the SM theory, called the Standard Model Extension (SME), which takes into account the gravitational interaction and all the factors that violate the CPT and Lorentz symmetries, by background tensor fields \cite{kosteleckycpt,kosteleckyviolation}. The background fields that appear in the SME may originate from phase transitions in a more fundamental theory that deals with the unification between quantum field theory and general relativity. Thus, the search for spontaneous Lorentz symmetry violation lies in the fact that this process can be an observable vestige of a primordial theory. This model is expected to describe the nature on a Planck energy scale, i.e. $10^{19}$ GeV \cite{violacao}. 

The SME takes into account the symmetry breaking in all sectors of the SM. The present work will only discuss what concerns electromagnetic interactions. Focusing in particular on Compton scattering \cite{compton,diagramas}. There are some experiments and experimental data obtained for Compton scattering, such as \cite{experimental1,experimental2,experimental3,experimental4}. However, these results are achieved in reactions at very low temperatures, close to zero. Although studies at finite temperatures are interesting, there are few investigations in this context \cite{comptonmatsubara}. Therefore, it is relevant to analyze this process subject to a finite temperature taking into account situations of the early universe. Processes such as Compton scattering at finite temperatures are important in situations where the temperatures involved are of cosmological order, such as inside large celestial bodies. Then it is necessary to look for formalisms that introduce temperature effects in the usual quantum field theories at zero temperature. For this, formalisms based on real-time or imaginary-time are used \cite{finite1,finite2,finite3}. Here a real-time formalism, called Thermofield Dynamics (TFD) formalism \cite{tfd1,lietfd}, is considered. This is a theory that introduces the temperature parameter by doubling the Hilbert space, based on the unitary transformation of operators \cite{realandimaginary,khannatfd}. With these ingredients, this work aims to obtain the differential cross section for Compton scattering at finite temperature, using the TFD formalism, and subject to Lorentz and CPT violation, by a constant background vector. This constant four-vector that leads to Lorentz violation can be time-like or space-like. Here, for simplicity, only the time-like case is chosen. 

This paper is organized as follows. In section II, it is seen how to introduce thermal effects in a quantum field theory by TFD formalism and the thermal fermion propagator is obtained. The propagator is divided into two parts: temperature dependent and independent parts. In section III, extended quantum electrodynamics and its implications for the Compton scattering process are presented. The Feynman diagrams that describe this reaction are shown, as well as the modifications in their vertices due to the presence of the constant background field in the Lagrangian.  In section IV, the differential cross section is calculated and the results are discussed. In section V, some concluding remarks are presented. It should be noted that the natural units of particle physics, i.e. $\hbar=c=1$, are considered. Furthermore, the \textit{FeynCalc} package of the \textit{Mathematica} software is used for the calculation of scalar products and tensor contractions \cite{feyncalc}.

\section{Thermofield Dynamics}

This section is divided into two parts. In the first part, an introduction to the TFD formalism is presented. While in the second part, as an example, the Feynman propagator for the Dirac field at finite temperature is calculated. This thermal propagator will be used to calculate the probability amplitude for the Compton scattering process.

The TFD formalism is characterized by the duplication of the Hilbert space $\mathbb{S}$ by an identical copy $\widetilde{\mathbb{S}}$,  such that the thermal space is defined by $\mathbb{S}_T=\mathbb{S}\otimes\widetilde{\mathbb{S}}$. It is important to note that the non-tilde space has the usual meaning, that is, the entire observable quantity is in it, while the tilde space exists only for symmetry purposes and is not related to physical experiments. To obtain the operator in this new tilde space, the tilde-conjugation rules  are used, which for a dynamical operator $O$ are given by
\begin{equation}
        (O_iO_j)^{\widetilde{}}=\widetilde{O}_i\widetilde{O}_j;\quad\quad
        (cO_i+O_j)^{\widetilde{}}=c^{*}\widetilde{O}_i+\widetilde{O}_j;\quad\quad
        (O_i^{\dagger})^{\widetilde{}}=(\widetilde{O}_i)^{\dagger};\quad\quad
        (\widetilde{O}_i)^{\widetilde{}}=\pm O_i;
\end{equation}
where the sign $\pm$ is characteristic of quantization, being negative for fermions and positive for bosons and $c$ is a complex number. Furthermore, these spaces obey the following Lie algebra
\begin{equation}
    \begin{split}
    [O_j,O_k]=iC_{jk}^{l}O_l;\quad\quad
    [\widetilde{O}_j,\widetilde{O}_k]=-iC_{jk}^{l}\widetilde{O}_l;\quad\quad
    [\widetilde{O}_j,O_k]=0,
\end{split}
\end{equation}
with $C^{l}_{jk}$ being the Lie structure constants.

In TFD, there is a vector $\ket{0(\beta)}$, called thermal vacuum state, which for an observable $O$ gives its thermal average as
\begin{equation}
    \langle O \rangle=\bra{0(\beta)}O\ket{0(\beta)},
\end{equation}
where $\beta=1/k_BT$, such that $k_B$ is the Boltzmann constant and $T$ is the temperature. From this, the doublet notation is defined
\begin{equation}
    O^{a}=\begin{pmatrix}
        O\\\widetilde{O}^\dagger
    \end{pmatrix}.\label{eq4}
\end{equation}
Here $a=1,2.$

This same notation is used to write quantum field operators. For the fermionic case, we have
\begin{equation}
    \psi^{a}(x)=\int \frac{d^3p}{(2\pi)^3}\frac{1}{\sqrt{2p_0}}\sum_s\left[a^{a}_s(p)u_s(p)e^{-ip\cdot x}+b^{\dagger a}_s(p)v_s(p)e^{ip\cdot x}\right],
\end{equation}
where $p_0$ is the energy,  $a^{a}_s(p)$ and $b^{\dagger a}_s(p)$ are annihilation and creation operators for particles and antiparticles, respectively,  and $u(p)$ and $v(p)$ are Dirac spinors.
In addition, in the tilde and non-tilde spaces, operators can be written in terms of those that lie in the thermal space using the Bogoliubov transformations.  Considering an arbitrary operator $O$, as described by the matrix form given in Eq. (\ref{eq4}), it is possible to write the transformation relation $ O^{a}=\mathbb{M}^{ab}(\beta)O_{b}(\beta)$, where $O_{b}(\beta)$ represents the same doublet operators in a thermal space. For fermions $\mathbb{M}^{ab}(\beta)$ is given as
\begin{equation}
   \mathbb{M}^{ab}(\beta)=\begin{pmatrix}
       U(\beta)&V(\beta)\\-V(\beta)&U(\beta)
   \end{pmatrix},\label{eq17}
\end{equation}
where $U^2=1-f$ and $V^2=f$, with $f$ being the Fermi-Dirac distribution. And for bosons
\begin{equation}
       \mathbb{M}^{ab}(\beta)=\begin{pmatrix}
       U^\prime(\beta)&V^\prime(\beta)\\V^\prime(\beta)&U^\prime(\beta)
   \end{pmatrix},\label{eq18}
\end{equation}
where $(U^\prime)^2=1+n$ and $(V^\prime)^2=n$, with $n$ being the Bose-Einstein distribution. It is important to note that the statistical distributions are defined as
\begin{equation}
    f(p_0)=\frac{1}{e^{\beta p_0}+1};\quad\quad n(k_0)=\frac{1}{e^{\beta k_0}-1}.
\end{equation}

Explicitly noting the creation and annihilation operators, in addition to Eqs. (\ref{eq4}), (\ref{eq17}) and (\ref{eq18}),  these transformations are written as
\begin{equation}
    \begin{split}
        a_s(p)&=U(\beta)a_s(\beta,p)+V(\beta)\widetilde{a}_s^\dagger(\beta,p);\quad\quad a_s^\dagger(p)=U(\beta)a_s^\dagger(\beta,p)+V(\beta)\widetilde{a}_s(\beta,p);\quad\quad\\\tilde{a}_s(p)&=U(\beta)\widetilde{a}_s(\beta,p)-V(\beta)a^\dagger_s(\beta,p);\quad\quad \widetilde{a}_s^\dagger(p)=U(\beta)\widetilde{a}^\dagger_s(\beta,p)-V(\beta)a_s(\beta,p);\quad\quad
    \end{split}\label{eq13}
\end{equation}
for fermions, as well as
\begin{equation}
    \begin{split}
        d_{\lambda}(k)&=U^{\prime}(\beta)d_{\lambda}(\beta,k)+V^{\prime}(\beta)\widetilde{d}^\dagger_{\lambda}(\beta,k);\quad\quad d_{\lambda}^\dagger(k)=U^{\prime}(\beta)d^\dagger_{\lambda}(\beta,k)+V^{\prime}(\beta)\widetilde{d}_{\lambda}(\beta,k);\quad\quad\\\tilde{d}_{\lambda}(k)&=U^{\prime}(\beta)\widetilde{d}_{\lambda}(\beta,k)+V^{\prime}(\beta)d^\dagger_{\lambda}(\beta,k);\quad\quad \widetilde{d}^\dagger_{\lambda}(k)=U^{\prime}(\beta)\widetilde{d}^\dagger_{\lambda}(\beta,k)+V^{\prime}(\beta)d_{\lambda}(\beta,k);\quad\quad
    \end{split}\label{eq14}
\end{equation}
for bosons, where $s$ and ${\lambda}$ are the spin and polarization indices, respectively.

In such a way that the thermal fermionic operators obey the following anticommutation relations
\begin{equation}
     \left\{a_s(\beta,p),a^\dagger_{s^{\prime}}(\beta,p^{\prime})\right\}=\left\{\tilde{a}_s(\beta,p),\tilde{a}^\dagger_{s^{\prime}}(\beta,p^{\prime})\right\}=(2\pi)^3\delta^3(\Vec{p}-\Vec{p}^{\prime})\delta_{s,s^{\prime}}.\label{eq15}
\end{equation}
The same is true for antiparticle operators $b$, $b^\dagger$, $\widetilde{b}$ and $\widetilde{b}^\dagger$. Regarding the bosonic operators, the commutation relations are given by
\begin{equation}
     \left[d_{\lambda}(\beta,k),d^\dagger_{{\lambda}^{\prime}}(\beta,k^{\prime})\right]=\left[\tilde{d}_{\lambda}(\beta,k),\tilde{d}^\dagger_{{\lambda}^{\prime}}(\beta,k^{\prime})\right]=(2\pi)^3\delta^3(\Vec{k}-\Vec{k}^{\prime})\delta_{{\lambda},{\lambda}^{\prime}}.\label{eq16}
\end{equation}
All other relations are equal to zero.

The meaning of Eqs. (\ref{eq13})-(\ref{eq16}) by the transformation matrices (\ref{eq17}) and (\ref{eq18}) is very simple. Operators with  dependence on $\beta$ create and annihilate particles characterized by temperature-dependents states in $\mathbb{S}_T$ space, just as the tilde and non-tilde operators create and annihilate particles in the spaces $\mathbb{S}$ and $\widetilde{\mathbb{S}}$, respectively, but at zero temperature.

Now,  in the context of the TFD formalism, let us analyze the propagator for the electron field. It is known that the Feynman propagator for the Dirac field is the probability amplitude of a fermion being created at some space-time point $x$ and propagates freely until it is annihilated at another point $y$. This quantity in the TFD formalism is given by
\begin{eqnarray}
        \Delta^{ab}(x-y)&=&-i\bra{0(\beta)}\mathbb{T}\left[\psi^a(x)\bar{\psi}^b(y)\right]\ket{0(\beta)}\label{eq08}\\
        &=&-i\left[\Theta(x^0-y^0)\bra{0(\beta)}\psi^{a}(x)\bar{\psi}^{b}(y)\ket{0(\beta)}-\Theta(y^0-x^0)\bra{0(\beta)}\bar{\psi}^{a}(y)\psi^b(x)\ket{0(\beta)}\right],\nonumber
\end{eqnarray}
where $\mathbb{T}$ is the time ordering operator, $\Theta(x)$ is the Heaviside step-function and $a,b=1,2$ are the thermal indices that characterize the duplicated element, implying that the propagator is a $2\times2$ matrix. In terms of the Fourier transform, the propagator is written as
\begin{equation}
    \Delta^{ab}(x-y)=\int \frac{d^4p}{(2\pi)^4}e^{-ip(x-y)}\Delta^{ab}(p),
\end{equation}
with $\Delta^{ab}(p)$ being the $ab$ element of the matrix representing the propagator in momentum space. This, in turn, is rewritten as
\begin{equation}
   \Delta^{ab}(p)=\int d^4z e^{ipz})\Delta^{ab}(z)\label{eq09}
\end{equation}
with $z=x-y$. Substituting Eq. (\ref{eq08}) in Eq. (\ref{eq09}) and taking the element $a=b=1$, we get
\begin{equation}
    \Delta(p)=-\int d^4ze^{ipz}i\left[\Theta(z^0)\bra{0(\beta)}\psi(x)\bar{\psi}(y)\ket{0(\beta)}-\Theta(-z^0)\bra{0(\beta)}\bar{\psi}(y)\psi(x)\ket{0(\beta)}\right],\label{eq10}
\end{equation}
where $\psi^1=\psi$ and $\Delta^{11}(p)=\Delta(p)$ have been used.
These averages are given as
\begin{eqnarray}
       \bra{0(\beta)}\psi(x)\bar{\psi}(y)\ket{0(\beta)}&=&\int \frac{d^3p}{(2\pi)^3}\frac{d^3p^{\prime}}{(2\pi)^3}\frac{1}{\sqrt{2p_0}\sqrt{2p^\prime_0}}\nonumber\\
       &&\times\sum_{s,s^\prime}\left[u_s(p)\bar{u}_{s^\prime}(p^\prime)e^{-i(px-p^\prime y)}\bra{0(\beta)}a_s(p)a^\dagger_{s^\prime}(p^\prime)\ket{0(\beta)}\right.\nonumber\\
       &&+\left.v_s(p)\bar{v}_{s^\prime}(p^\prime)e^{i(px-p^\prime y)}\bra{0(\beta)}b^\dagger_s(p)b_{s^\prime}(p^\prime)\ket{0(\beta)}\right]\label{17}
\end{eqnarray}
and
\begin{eqnarray}
         \bra{0(\beta)}\bar{\psi}(y)\psi(x)\ket{0(\beta)}&=&\int \frac{d^3p}{(2\pi)^3}\frac{d^3p^{\prime}}{(2\pi)^3}\frac{1}{\sqrt{2p_0}\sqrt{2p^\prime_0}}\nonumber\\
         &&\times\sum_{s,s^\prime}\left[u_s(p)\bar{u}_{s^\prime}(p^\prime)e^{i(p^\prime y-px)}\bra{0(\beta)}a^\dagger_{s^\prime}(p^\prime)a_s(p)\ket{0(\beta)}\right.\nonumber\\
         &&+\left.v_s(p)\bar{v}_{s^\prime}(p^\prime)e^{-i(p^\prime y-px)}\bra{0(\beta)}b_{s^\prime}(p^\prime)b^\dagger_s(p)\ket{0(\beta)}\right].\label{18}
\end{eqnarray}
Using the Bogoliubov transformations Eq. (\ref{eq13}) and the anticommutation relations Eq. (\ref{eq15}), we get
\begin{gather}
    \bra{0(\beta)}a_s(p)a^\dagger_{s^\prime}(p^\prime)\ket{0(\beta)}=U^2\bra{0(\beta)}a_s(\beta,p)a^\dagger_{s^\prime}(\beta,p^\prime)\ket{0(\beta)}=U^2(2\pi)^3\delta^3(\Vec{p}-\Vec{p}^{\prime})\delta_{s,s^{\prime}};\\
    \bra{0(\beta)}b^\dagger_s(p)b_{s^\prime}(p^\prime)\ket{0(\beta)}=V^2\bra{0(\beta)}\tilde{b}_s(\beta,p)\tilde{b}^\dagger_{s^\prime}(\beta,p^\prime)\ket{0(\beta)}=V^2(2\pi)^3\delta^3(\Vec{p}-\Vec{p}^{\prime})\delta_{s,s^{\prime}};\\
    \bra{0(\beta)}a^\dagger_{s^\prime}(p^\prime)a_s(p)\ket{0(\beta)}=V^2\bra{0(\beta)}\tilde{a}_s(\beta,p)\tilde{a}^\dagger_{s^\prime}(\beta,p^\prime)\ket{0(\beta)}=V^2(2\pi)^3\delta^3(\Vec{p}-\Vec{p}^{\prime})\delta_{s,s^{\prime}};\\
    \bra{0(\beta)}b_{s^\prime}(p^\prime)b^\dagger_s(p)\ket{0(\beta)}=U^2\bra{0(\beta)}b_s(\beta,p)b^\dagger_{s^\prime}(\beta,p^\prime)\ket{0(\beta)}=U^2(2\pi)^3\delta^3(\Vec{p}-\Vec{p}^{\prime})\delta_{s,s^{\prime}}.
\end{gather}
Then Eqs. (\ref{17}) and (\ref{18}) become
\begin{eqnarray}
        \bra{0(\beta)}\psi(x)\bar{\psi}(y)\ket{0(\beta)}&=&\int \frac{d^3p}{(2\pi)^3}\frac{1}{2p_0}\sum_s\left[U^2u_s(p)\bar{u}_{s}(p)e^{-ip(x-y)}+V^2v_s(p)\bar{v}_s(p)e^{ip(x-y)}\right]\nonumber\\
        &=&\int \frac{d^3p}{(2\pi)^3}\frac{1}{2\xi}\left[U^2(\gamma^0\xi-\gamma^jp_j+m)e^{-i\xi z^0}e^{ip^jz_j}\right.\nonumber\\
        &&+\left.V^2(\gamma^0\xi-\gamma^jp_j-m)e^{i\xi z^0}e^{-ip^jz_j}\right]\label{eq11}
\end{eqnarray}
and
\begin{eqnarray}
        \bra{0(\beta)}\bar{\psi}(y)\psi(x)\ket{0(\beta)}&=&\int \frac{d^3p}{(2\pi)^3}\frac{1}{2p_0}\sum_s\left[V^2u_s(p)\bar{u}_{s}(p)e^{-ip(x-y)}+U^2v_s(p)\bar{v}_s(p)e^{ip(x-y)}\right]\nonumber\\
        &=&\int \frac{d^3p}{(2\pi)^3}\frac{1}{2\xi}\left[V^2(\gamma^0\xi-\gamma^jp_j+m)e^{-i\xi z^0}e^{ip^jz_j}\right.\nonumber\\
        &&\left.+U^2(\gamma^0\xi-\gamma^jp_j-m)e^{i\xi z^0}e^{-ip^jz_j}\right].\label{eq12}
\end{eqnarray}
Where the completeness relations  have been used, i.e.,
\begin{equation}
    \sum_{s}u_s(p)\bar{u}_s(p)=(\gamma^0\xi-\gamma^jp_j+ m);\quad\quad \sum_{s}v_s(p)\bar{v}_s(p)=(\gamma^0\xi-\gamma^jp_j- m),
\end{equation}
with $\xi^2=\vec{p}^2+m^2$.

Putting Eqs. (\ref{eq11}) and (\ref{eq12}) into Eq. (\ref{eq10}) results in
\begin{eqnarray}
       \Delta(p^\prime)&=&-\int\frac{d^3p}{(2\pi)^3}\frac{1}{2\xi} \int d^4ze^{ip^{\prime}z}i\Bigl\{\Theta(z^0)\Bigl[U^2(\gamma^0\xi-\gamma^jp_j+m)e^{-i\xi z^0}e^{ip^jz_j}\nonumber\\
        &&+V^2(\gamma^0\xi-\gamma^jp_j-m)e^{i\xi z^0}e^{-ip^jz_j}\Bigl]\nonumber\\
        &&-\Theta(-z^0)\left[V^2(\gamma^0\xi-\gamma^jp_j+m)e^{-i\xi z^0}e^{ip^jz_j}+U^2(\gamma^0\xi-\gamma^jp_j-m)e^{i\xi z^0}e^{-ip^jz_j}\right]\Bigl\}\nonumber\\
        &=&-\int d^3p\frac{1}{2\xi}\int dz^0e^{ip^\prime_0z^0}i\Bigl\{\Theta(z^0)\Bigl[U^2(\gamma^0\xi-\gamma^jp_j+m)e^{-i\xi z^0}\delta^3(p-p^\prime)\\
        &&+V^2(\gamma^0\xi-\gamma^jp_j-m)e^{i\xi z^0}\delta^3(p+p^\prime)\Bigl]\nonumber\\
        &&-\Theta(-z^0)\left[V^2(\gamma^0\xi-\gamma^jp_j+m)e^{-i\xi z^0}\delta^3(p-p^\prime)+U^2(\gamma^0\xi-\gamma^jp_j-m)e^{i\xi z^0}\delta^3(p+p^\prime)\right]\Bigl\}.\nonumber
\end{eqnarray}
Performing the integration in $p$ leads to
\begin{eqnarray}
       \Delta(p)&=&-\int_0^{\infty}dz^0i\left[U^2\frac{(\gamma^0\xi-\gamma^jp_j+m)}{2\xi}e^{-i(\xi-p_0) z^0}+V^2\frac{(\gamma^0\xi+\gamma^jp_j-m)}{2\xi}e^{i(\xi+p_0) z^0}\right]\nonumber\\
        &&+\int^0_{-\infty}dz^0i\left[V^2\frac{(\gamma^0\xi-\gamma^jp_j+m)}{2\xi}e^{-i(\xi-p_0) z^0}+U^2\frac{(\gamma^0\xi+\gamma^jp_j-m)}{2\xi}e^{i(\xi+p_0) z^0}\right].\label{27}
\end{eqnarray}

Looking at the $z^0$ integration, it can be seen that
\begin{equation}
\lim_{\eta\to0^{+}}\int_{0}^{\infty} dz^0 e^{-i(\xi- p_0-i\eta)z^0}=\lim_{\eta\to0^{+}}\frac{i}{p_0-\xi+i\eta},
\end{equation}
and
\begin{equation}
\lim_{\eta\to0^{+}}\int_{-\infty}^{0} dz^0 e^{i(\xi+ p_0-i\eta)z^0}=-\lim_{\eta\to0^{+}}\frac{i}{p_0+\xi-i\eta}.
\end{equation}
Then Eq. (\ref{27}) becomes
\begin{eqnarray}
       \Delta(p)&=&\lim_{\eta\to0^{+}}\left[U^2\frac{(\gamma^0\xi-\gamma^jp_j+m)}{2\xi}\frac{1}{p_0-\xi+i\eta}+V^2\frac{(\gamma^0\xi+\gamma^jp_j-m)}{2\xi}\frac{1}{p_0+\xi+i\eta}\right]\nonumber\\
        &&+\lim_{\eta\to0^{+}}\left[V^2\frac{(\gamma^0\xi-\gamma^jp_j+m)}{2\xi}\frac{1}{p_0-\xi-i\eta}+U^2\frac{(\gamma^0\xi+\gamma^jp_j-m)}{2\xi}\frac{1}{p_0+\xi-i\eta}\right].
\end{eqnarray}

Following the same procedure for the other elements of $\Delta^{ab}$, the complete matrix is written as
\begin{eqnarray}
       \Delta^{ab}(p)&=&\lim_{\eta\to0^{+}}\left(\frac{\gamma^0\xi-\gamma^jp_j+m}{2\xi}\right)\left[\mathbb{M}(p_0)(p_0-\xi+i\eta\tau)^{-1}\mathbb{M}^{\dagger}(p_0)\right]^{ab}\nonumber\\
        &&+\lim_{\eta\to0^{+}}\left(\frac{\gamma^0\xi+\gamma^jp_j-m}{2\xi}\right)\left[\mathbb{M}(-p_0)(p_0+\xi+i\eta\tau)^{-1}\mathbb{M}^{\dagger}(-p_0)\right]^{ab},\label{eq19}
\end{eqnarray}
where $\mathbb{M}$ is given by Eq. (\ref{eq17}), and
\begin{equation}
    \tau=\begin{pmatrix}
       1 & 0\\0 & -1
    \end{pmatrix}.
\end{equation}

For the study developed here, it is interesting to separate the propagator into temperature-independent and temperature-dependent parts. To achieve this goal, let us write
{\small
\begin{eqnarray}
        \mathbb{M}(p_0)(p_0-\xi+i\eta\tau)^{-1}\mathbb{M}^{\dagger}(p_0)&=&
    \begin{pmatrix}
       \frac{U^2}{p_0-\xi+i\eta} +\frac{V^2}{p_0-\xi-i\eta} & -\frac{UV}{p_0-\xi+i\eta} + \frac{VU}{p_0-\xi-i\eta}\\-\frac{UV}{p_0-\xi+i\eta} + \frac{VU}{p_0-\xi-i\eta} &  \frac{V^2}{p_0-\xi+i\eta} +\frac{U^2}{p_0-\xi-i\eta}
    \end{pmatrix}\nonumber\\
    &=&\begin{pmatrix}
        \frac{1}{p_0-\xi+i\eta} & 0\\0 & \frac{1}{p_0-\xi-i\eta}
    \end{pmatrix}\label{33}\\
    &&+  f(p_0)\begin{pmatrix}
        -\frac{1}{p_0-\xi+i\eta}+\frac{1}{p_0-\xi-i\eta} & -\frac{e^{\beta p_0/2}}{p_0-\xi+i\eta} + \frac{e^{\beta p_0/2}}{p_0-\xi-i\eta}\\-\frac{e^{\beta p_0/2}}{p_0-\xi+i\eta} + \frac{e^{\beta p_0/2}}{p_0-\xi-i\eta} &  \frac{1}{p_0-\xi+i\eta} -\frac{1}{p_0-\xi-i\eta}
    \end{pmatrix}.\nonumber
\end{eqnarray}
}
To avoid divergences, it is necessary to use some cutting rules \cite{cut1,cut2} which are given as
\begin{equation}
    \frac{1}{p_0-\xi+i\eta}=\frac{1}{p_0-\xi}-\pi i\delta(p_0-\xi);\quad\quad \frac{1}{p_0-\xi-i\eta}=\frac{1}{p_0-\xi}+\pi i\delta(p_0-\xi).
\end{equation}

In this way, Eq. (\ref{33}) reads
\begin{eqnarray}
  \mathbb{M}(p_0)(p_0-\xi+i\eta\tau)^{-1}\mathbb{M}^{\dagger}(p_0)&=&  \begin{pmatrix}
        \frac{1}{p_0-\xi+i\eta} & 0\\0 & \frac{1}{p_0-\xi-i\eta}
    \end{pmatrix}\nonumber\\
    &&+2\pi i\delta(p_0-\xi)f(p_0)\begin{pmatrix}
         1& e^{\beta p_0/2}\\e^{\beta p_0/2} &  -1
    \end{pmatrix}.\label{eq20}
\end{eqnarray}
Analogously, using the identity $U(-p_0)=V(p_0)$, we get
\begin{eqnarray}
        \mathbb{M}(-p_0)(p_0+\xi+i\eta\tau)^{-1}\mathbb{M}^{\dagger}(-p_0)&=&\begin{pmatrix}
            \frac{1}{p_0+\xi+i\eta}&0\\0&\frac{1}{p_0+\xi-i\eta}
        \end{pmatrix}\nonumber\\
        &&+2\pi i\delta(p_0+\xi)f(p_0)\begin{pmatrix}
         -1& e^{\beta p_0/2}\\e^{\beta p_0/2} &  1
    \end{pmatrix}.\label{eq21}
\end{eqnarray}

Here, it is observed that
{\small
\begin{eqnarray}
   && \left(\frac{\gamma^0\xi-\gamma^jp_j+m}{2\xi}\right)\begin{pmatrix}
        \frac{1}{p_0-\xi+i\eta} & 0\\0 & \frac{1}{p_0-\xi-i\eta}
    \end{pmatrix}+\left(\frac{\gamma^0\xi+\gamma^jp_j-m}{2\xi}\right)\begin{pmatrix}
            \frac{1}{p_0+\xi+i\eta}&0\\0&\frac{1}{p_0+\xi-i\eta}
        \end{pmatrix}\nonumber\\
        &&=\frac{\slashed{p}+m}{p^2-m^2}\label{eq22}
\end{eqnarray}
}
Then using Eqs. (\ref{eq20})-(\ref{eq22}) in Eq. (\ref{eq19}) the electron propagator, in the momentum space, is written as 
\begin{eqnarray}
\Delta^{ab}(p)=S^{(0)}(p)+S^{(\beta)}(p),
\end{eqnarray}
where
\begin{equation}
    S^{(0)}(p)=\frac{\slashed{p}+m}{p^2-m^2}\label{P0}
\end{equation}
is the zero temperature part, and
\begin{equation}
    S^{(\beta)}(p)=\frac{2\pi i}{(e^{\beta p_0}+1)}\left[\frac{(\gamma^0\xi-\gamma\cdot\Vec{p}+m)}{2\xi}\Delta_1\delta(p^0-\xi)+\frac{(\gamma^0\xi+\gamma\cdot\Vec{p}-m)}{2\xi}\Delta_2\delta(p^0+\xi)\right]\label{PT}
\end{equation}
is the temperature dependent part, as obtained in \cite{ale1}. The terms $\Delta_1$ and $\Delta_2$ are thermal matrices given by
\begin{equation}
    \Delta_1=\begin{pmatrix}
    1 & e^{\beta p_0/2}\\e^{\beta p_0/2}&-1\end{pmatrix};\quad\quad \Delta_2=\begin{pmatrix}
    -1 & e^{\beta p_0/2}\\e^{\beta p_0/2}&1\end{pmatrix}.
\end{equation}

It is important to note that even though the propagator is a $2\times2$ matrix, the elements represent mixtures between tilde and non-tilde spaces. Therefore, as only the non-tilde space is related to the observables, in the final result of any calculated quantity it is necessary to take only the element $a=b=1$ from the resulting matrix.

\section{Compton scattering}

In this section, a Lorentz violating theory is introduced and the probability amplitude for Compton scattering is calculated. To deal with extended quantum electrodynamics, which leads to the Lorentz violation, a constant background vector is introduced into the standard QED theory via covariant derivative \cite{ale2}. The Lagrangian that describes this scattering process on a Lorentz-violating background is given as
\begin{equation}
    \mathcal{L}=-\frac{1}{4}F_{\mu\nu}F^{\mu\nu}+\Bar{\psi}\left(iD_\mu\gamma^\mu-m\right)\psi,\label{eq01}
\end{equation}
where $D_\mu=\partial_\mu-ieA_\mu+igb^\nu{}^{*}F_{\mu\nu}$, such  that $^{*}F_{\mu\nu}=\frac{1}{2}\varepsilon_{\mu\nu\alpha\kappa}F^{\alpha\kappa}$ is the dual electromagnetic tensor, $g$ is the coupling constant and $b^\nu$ the Lorentz constant vector. It is important to point out that a new interaction is introduced via the covariant derivative, this is an alternative procedure to implement Lorentz violation. Furthermore, it is interesting to note that there are two different ways to introduce non-minimal coupling into the covariant derivative: (i) $igb^\nu{}^{*}F_{\mu\nu}$ and (ii) $igb^\nu{}F_{\mu\nu}$, where the difference consists in using the dual of the field strength or the field strength itself. As discussed in reference \cite{First}, the choice of one term or another implies different physical situations. For example, using the term with the dual of the field strength in the Dirac equation it is possible to study the induction of the Aharonov-Casher effect, while using the field strength itself this effect is not yielded, but leads to an extra phase involving the magnetic field. Here the first term is considered since the second term was already used in \cite{comptonviolation} to calculate the cross-section for Compton scattering at zero temperature. In addition, this new coupling involving the 5-dimensional operator leads to a nonrenormalizable theory at power counting. However, it is not a problem for the study developed here, since only the tree-level scattering processes are analyzed.  Starting from the fact that the photonic field can be written by a superposition of plane waves with a well-defined momentum $q$, that is, $A_\mu\propto e^{-iqx}$, the covariant derivative is written as
\begin{equation}
    D_{\mu}=\partial_\mu-ieA_\mu+g\varepsilon_{\mu\nu\alpha\kappa}b^\nu q^{\alpha}A^\kappa.
\end{equation}

\begin{figure}[ht]
    \centering
    \includegraphics[scale=0.2]{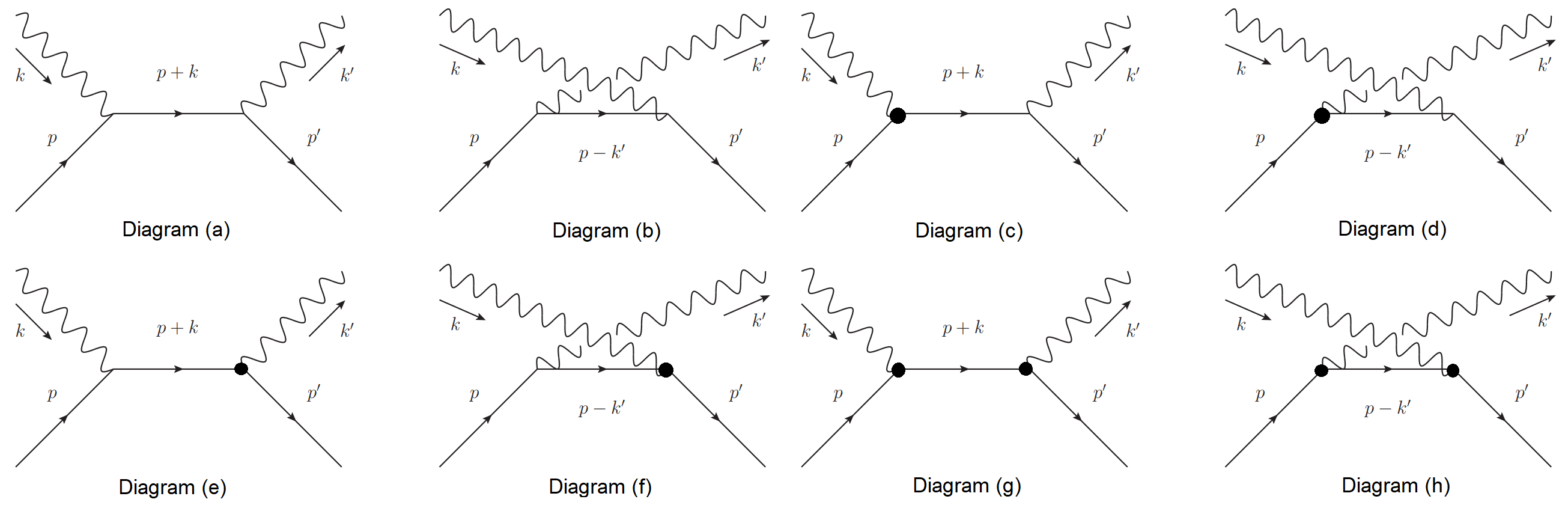}
    \caption{Feynman diagrams for the Lorentz violating Compton scattering. Adapted from \cite{compton}}
    \label{fig1}
\end{figure}

Then the Lagrangian (\ref{eq01}) is divided into two parts, one that contains only the free fields and another that expresses the interaction between them. The interaction Lagrangian is given by
\begin{equation}
    \mathcal{L}_{\text{int}}=e\Bar{\psi}\gamma^\mu A_\mu\psi+ig\Bar{\psi}\varepsilon_{\mu\nu\alpha\kappa}\gamma^\mu b^\nu q^\alpha A^\kappa\psi.
\end{equation}
This implies Feynman diagrams with two different types of vertices, those of the usual QED and those with Lorentz violation (characterized by a filled circle). All of them are shown in Figure \ref{fig1}. Note that although the additional vertex explicitly violates Lorentz symmetry, since the vector $b^\mu$ defines a privileged direction in space-time, it is gauge invariant.

The interaction Hamiltonian is
\begin{equation}
H_{\text{int}}=\int d^3x(-\mathcal{L}_{\text{int}})=\int d^3x\left(-e\Bar{\psi}\gamma^\mu A_\mu\psi-ig\Bar{\psi}\varepsilon_{\mu\nu\alpha\kappa}\gamma^\mu b^\nu q^\alpha A^\kappa\psi\right).
\end{equation}
From which the second order element of the thermal scattering matrix is written as
\begin{equation}
    \hat{S}^{(2)}=\frac{(-i)^2}{2!}\int dx^0dy^0:\hat{H}_{\text{int}}(x)\hat{H}_{\text{int}}(y):\;=\frac{(-i)^2}{2!}\int d^4xd^4y:\mathcal{L}_{\text{int}}(x)\mathcal{L}_{\text{int}}(y):,
\end{equation}
where only non-tilde operators are considered, since they are the true physical observables. The normal ordering operator ``$:$'' is used. It makes all creation and annihilation operators always act correctly, i.e., in order to keep only finite quantities and avoid divergences. 

The initial and final states that characterize this scattering process at finite temperature are given, respectively, by
\begin{equation}
\ket{i(\beta)}=a^{\dagger}_s(\beta,p)d^{\dagger}_\lambda(\beta,k)\ket{0(\beta)};\quad\quad\ket{f(\beta)}=a^{\dagger}_{s^\prime}(\beta,p^{\prime})d^{\dagger}_{\lambda^\prime}(\beta,k^{\prime})\ket{0(\beta)},
\end{equation}
such that the probability amplitude $\mathcal{M}$ for this reaction to occur is 
\begin{equation}
    \mathcal{M}=\bra{f(\beta)}\hat{S}^{(2)}\ket{i(\beta)}=\mathcal{M}_0+\mathcal{M}_1+\mathcal{M}_2
\end{equation}
where $\mathcal{M}_0$, $\mathcal{M}_1$ and $\mathcal{M}_2$ represent the events without one, only one and with two violation vertices, respectively.

Using Feynman diagrams, analyzing the operators that contribute to the scattering, one can write
\begin{eqnarray}
    i\mathcal{M}_0&=&-e^2\int d^4xd^4y\bra{f(\beta)}:\Bar{\psi}^{-}_x\gamma^\mu i[\psi_x\Bar{\psi}_y]\gamma^\nu\psi^{+}_y(A_\mu^{-})_x(A_\nu^{+})_y:\ket{i(\beta)}\nonumber\\
   &&-e^2\int d^4xd^4y\bra{f(\beta)}:\Bar{\psi}^{-}_x\gamma^\mu i[\psi_x\Bar{\psi}_y]\gamma^\nu\psi^{+}_y(A_\mu^{+})_x(A_\nu^{-})_y:\ket{i(\beta)}\nonumber\\
    &\equiv& i\mathcal{M}_0^{(a)}+i\mathcal{M}_0^{(b)}.
\end{eqnarray}
Analogously for diagrams with a violation vertex
\begin{eqnarray}
        i\mathcal{M}_1&=&2igb^\nu e\varepsilon_{\mu\nu\alpha\kappa}\int d^4xd^4y\bra{f(\beta)}:\Bar{\psi}^{-}_x\gamma^\sigma i[\psi_x\Bar{\psi}_y]\gamma^\mu\psi^{+}_y(A_\sigma^{-})_x(k^{\alpha}A^{\kappa+})_y:\ket{i(\beta)}\nonumber\\
       & &+2igb^\nu e\varepsilon_{\mu\nu\alpha\kappa}\int d^4xd^4y\bra{f(\beta)}:\Bar{\psi}^{-}_x\gamma^\sigma i[\psi_x\Bar{\psi}_y]\gamma^\mu\psi^{+}_y(A_\sigma^{+})_x(k^{\alpha}A^{\kappa-})_y:\ket{i(\beta)}\nonumber\\
       &\equiv& i\mathcal{M}_1^{(a)}+i\mathcal{M}_1^{(b)}.
\end{eqnarray}
Note that the contribution of diagrams (c) and (d) given by Figure \ref{fig1} is the same of the (e) and (f), this fact is reinforced by a factor of 2 in $\mathcal{M}_1$. Finally, for diagrams with two violation vertices
\begin{eqnarray}
        i\mathcal{M}_2&=&g^2\varepsilon_{\mu\nu\alpha\kappa}\varepsilon_{\omega\rho\sigma\tau}b^\nu b^\rho\int d^4xd^4y\bra{f(\beta)}:\Bar{\psi}^{-}_x\gamma^\mu i[\psi_x\Bar{\psi}_y]\gamma^\omega\psi^{+}_y(k^\alpha A^{\kappa-})_x(k^{\sigma}A^{\tau+})_y:\ket{i(\beta)}\nonumber\\
        &&+g^2\varepsilon_{\mu\nu\alpha\kappa}\varepsilon_{\omega\rho\sigma\tau}b^\nu b^\rho\int d^4xd^4y\bra{f(\beta)}:\Bar{\psi}^{-}_x\gamma^\mu i[\psi_x\Bar{\psi}_y]\gamma^\omega\psi^{+}_y(k^\alpha A^{\kappa+})_x(k^{\sigma}A^{\tau-})_y:\ket{i(\beta)}\nonumber\\
        &\equiv& i\mathcal{M}_2^{(a)}+i\mathcal{M}_2^{(b)}.
\end{eqnarray}

For each contribution, there are two diagrams, one for the annihilation channel (whose momentum conservation is characterized by $p+k$) and one for the exchange channel (charaterized by $p-k^\prime$). Considering the integration and the creation and annihilation operators, we obtain
\begin{eqnarray}
   i\mathcal{M}_0 &=&-e^2F(\beta) \Bigl[\bar{u}(p^{\prime})\gamma^\mu i\Delta(p+k)\gamma^\nu u(p)\epsilon_\mu^{\dagger}(k^{\prime})\epsilon_\nu(k)\nonumber\\
   &&+\bar{u}(p^{\prime})\gamma^\mu i\Delta(p-k^{\prime})\gamma^\nu u(p)\epsilon_\mu(k)\epsilon_\nu^{\dagger}(k^{\prime})\Bigl].\label{eq02}
\end{eqnarray}
In the same way we get
\begin{eqnarray}
        i\mathcal{M}_1&=&2iegF(\beta) \varepsilon_{\mu\nu\alpha\kappa} b^\nu\Bigl[\bar{u}(p^{\prime})\gamma^\sigma i\Delta(p+k)\gamma^\mu u(p)k^{\alpha}\epsilon^{\dagger}_\sigma(k^{\prime})\epsilon^{\kappa}(k)\nonumber\\
        &&+\bar{u}(p^{\prime})\gamma^\sigma i\Delta(p-k^{\prime})\gamma^\mu u(p)\epsilon_\sigma(k)(k^{\prime})^{\alpha}\epsilon^{\dagger\kappa}(k^{\prime})\Bigl]\label{eq03}
\end{eqnarray}
and
\begin{eqnarray}
       i\mathcal{M}_2&=& g^2F(\beta)\varepsilon_{\mu\nu\alpha\kappa}\varepsilon_{\omega\rho\sigma\tau}b^\nu b^\rho \Bigl[\bar{u}(p^{\prime})\gamma^\mu i\Delta(p+k)\gamma^\omega u(p)k^{\sigma}(k^{\prime})^{\alpha}\epsilon^{\dagger\kappa}(k^{\prime})\epsilon^{\tau}(k)\nonumber\\
&&+\bar{u}(p^{\prime})\gamma^\mu i\Delta(p-k^{\prime})\gamma^\omega u(p)(k^{\prime})^{\sigma}k^{\alpha}\epsilon^{\kappa}(k)\epsilon^{\dagger\tau}(k^{\prime})\Bigl].\label{eq04}
\end{eqnarray}
Here the temperature-dependent proportionality function is defined as 
\begin{equation}
    F^2(\beta)=[1+n(\omega_i)][1+n(\omega_f)][1-f(E_i)][1-f(E_f)].
\end{equation}
It is important to note that a similar result, for the function $F^2(\beta)$, was obtained by \cite{comptonmatsubara} using another thermal formalism. 

\begin{figure}[ht]
    \centering
    \includegraphics[scale=0.4]{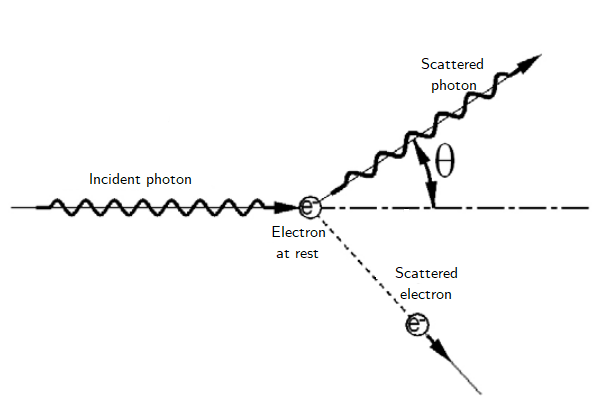}
    \caption{Illustration of the scattering process in the laboratory frame. Here the $z$ axis is horizontal, while the $x$ axis is in the vertical direction. Adapted from \cite{comptonfig}.}
    \label{fig2}
\end{figure}

Taking the laboratory frame as shown in Figure \ref{fig2}, the incoming and outgoing momenta of electrons and photons are written as
\begin{eqnarray}
    p&=&(m,0,0,0);\quad\quad\quad\quad\quad\quad\quad\quad\quad\quad k=(\omega,0,0,\omega);\quad\quad \nonumber\\
    p^{\prime}&=&(E,-\omega^\prime\sin{\theta},0,\omega-\omega^\prime\cos{\theta});\quad\quad k^{\prime}=(\omega^\prime,\omega^\prime\sin{\theta},0,\omega^\prime\cos{\theta}),
\end{eqnarray}
where $E=m+\omega-\omega^{\prime}$ due to energy conservation.

For the annihilation channel, that is, for the diagrams (a), (c), (e) and (g) of Figure \ref{fig1}, looking at Eqs. (\ref{P0}) and (\ref{PT}) the propagator takes the form
\begin{eqnarray}
S(p+k)&=&\frac{\slashed{p}+\slashed{k}+m}{(p+k)^2-m^2}+\frac{\pi i}{\xi_a[e^{\beta( m+\omega)}+1]}\Bigl[\left(\gamma^0\xi_a-\gamma^3\omega+m\right)\Delta_1\delta(m+\omega-\xi_a)\nonumber\\
&&+\left(\gamma^0\xi_a+\gamma^3\omega-m\right)\Delta_2\delta(m+\omega+\xi_a)\Bigl],
\end{eqnarray}
and for the exchange channel, described by the other diagrams, we have
\begin{eqnarray}
        S(p-k^{\prime})&=&\frac{\slashed{p}-\slashed{k}^\prime+m}{(p-k^\prime)^2-m^2}+\frac{\pi i}{\xi_b[e^{\beta (m-\omega^\prime)}+1]}\Bigl[(\gamma^0\xi_b-\gamma^j(p-k^{\prime})_j+m)\Delta_1\delta(m-\omega^\prime-\xi_b)\nonumber\\
        &&+(\gamma^0\xi_b+\gamma^j(p-k^{\prime})_j-m)\Delta_2\delta(m-\omega^\prime+\xi_b)\Bigl],
\end{eqnarray}
where $\xi_a=\sqrt{\omega^2+m^2}$ and $\xi_b=\sqrt{(p-k^{\prime})_j^2+m^2}$.

From the total probability amplitude $\mathcal{M}$, the total probability density $|\mathcal{M}|^2$ is obtained. Consequently, for unpolarized beams, their average over all possible spin and polarization states becomes
\begin{eqnarray}
        \langle |\mathcal{M}|^2\rangle&=&\langle |\mathcal{M}_0+\mathcal{M}_1+\mathcal{M}_2|^2\rangle\nonumber\\
        &=&\langle|\mathcal{M}_0|^2+\mathcal{M}_0\mathcal{M}_1^\dagger+\mathcal{M}_1\mathcal{M}_0^\dagger+|\mathcal{M}_1|^2+\mathcal{M}_0\mathcal{M}_2^\dagger+\mathcal{M}_2\mathcal{M}_0^\dagger+\mathcal{O}(g^3)\rangle\nonumber\\
&\approx&\langle|\mathcal{M}_0|^2\rangle +\langle \mathcal{M}_0\mathcal{M}_1^\dagger\rangle +\langle \mathcal{M}_1\mathcal{M}_0^\dagger\rangle +\langle|\mathcal{M}_1|^2\rangle +\langle \mathcal{M}_0\mathcal{M}_2^\dagger\rangle +\langle \mathcal{M}_2\mathcal{M}_0^\dagger\rangle,    
\label{eq001}
\end{eqnarray}
where terms with a dependence greater than second order on the parameter that violates Lorentz symmetries have been ignored.

Notice that from Eqs. (\ref{eq02}) - (\ref{eq04}), $\mathcal{M}_0$ and $\mathcal{M}_2$ imply a real value, while $\mathcal{M}_1$ in pure imaginary, therefore $\mathcal{M}_0\mathcal{M}_1^{\dagger}=-\mathcal{M}_1\mathcal{M}_0^\dagger$, as well as $\mathcal{M}_0\mathcal{M}_2^\dagger=\mathcal{M}_2\mathcal{M}_0^\dagger$. Thus, from Eq. (\ref{eq001}), in a second-order approximation to the coupling constant $g$, it is written
\begin{equation}
     \langle |\mathcal{M}|^2\rangle=\langle|\mathcal{M}_0|^2\rangle+\langle|\mathcal{M}_1|^2\rangle +2\langle \mathcal{M}_0\mathcal{M}_2^\dagger\rangle.\label{eq002}
\end{equation}

All the following results are presented in terms of the Mandelstam variables in the Lab frame, that is,
\begin{equation}
    s=2m\omega+m^2;\quad\quad t=2\omega\omega^\prime(\cos{\theta}-1);\quad\quad u=-2m\omega^\prime+m^2,
\end{equation}
where $\theta$ is the scattering angle.

Now term by term of Eq. (\ref{eq002}) is calculated separately. The average takes into account the amount of  initial and final particles for each reaction, as well as a sum over all possible spin and polarization states, so that the first term of Eq. (\ref{eq002}) is written as
\begin{equation}
    \langle|\mathcal{M}_0|^2\rangle=\frac{1}{4}\sum_{\substack{\lambda,\lambda^{\prime}\\s,s^{\prime}}}|\mathcal{M}_0|^2=\langle|\mathcal{M}_0^{(a)}|^2\rangle+\langle|\mathcal{M}_0^{(b)}|^2\rangle+\left\langle2\Re{\mathcal{M}_0^{(a)\dagger}\mathcal{M}_0^{(b)}}\right\rangle.\label{eq24}
\end{equation}
Evaluating all diagrams, taking the completeness relations
\begin{equation}
\sum_{s}u(p,s)\bar{u}(p,s)=\slashed{p}+m;\quad\quad\sum_{\lambda}\epsilon_\mu(k,\lambda)\epsilon^{\dagger}_\alpha(k,\lambda)=g_{\mu\alpha},
\end{equation}
and the relation
\begin{equation}
\sum_{s,s^\prime}[\bar{u}(p,s)\gamma^\mu\gamma^\nu u(p^\prime,s^\prime)\bar{u}(p^\prime,s^\prime)\gamma_\mu\gamma_\nu u(p,s)]=\Tr\sum_{s,s^\prime}[u(p,s)\bar{u}(p,s)\gamma^\mu\gamma^\nu u(p^\prime,s^\prime)\bar{u}(p^\prime,s^\prime)\gamma_\mu\gamma_\nu],
\end{equation}
we get for the diagrams without violation vertex
\begin{equation}
\langle|\mathcal{M}_0^{(a)}|^2\rangle=\frac{e^4}{4}F^2(\beta)\Tr{(\slashed{p}^{\prime}+m)\gamma^\mu S(p+k)\gamma^\nu(\slashed{p}+m)\gamma_\nu S^\dagger(p+k)\gamma_\mu},
\end{equation}
\begin{equation}
\langle|\mathcal{M}_0^{(b)}|^2\rangle=\frac{e^4}{4}F^2(\beta)\Tr{(\slashed{p}^{\prime}+m)\gamma^\mu S(p-k')\gamma^\nu(\slashed{p}+m)\gamma_\nu S^\dagger(p-k')\gamma_\mu}
\end{equation}
and
\begin{equation}
   \left\langle2\Re{\mathcal{M}_0^{(a)\dagger}\mathcal{M}_0^{(b)}}\right\rangle=\frac{e^4}{2}F^2(\beta)\Re{\Tr[(\slashed{p}+m)\gamma_\mu S^{\dagger}(p+k)\gamma_\nu(\slashed{p}^{\prime}+m)\gamma^\mu S(p-k^\prime)\gamma^\nu]}.
\end{equation}

For the second term of Eq. (\ref{eq002}), using $\mathcal{M}_1=\mathcal{M}_1^{(a)}+\mathcal{M}_1^{(b)}$, is written
\begin{equation}
\langle|\mathcal{M}_1|^2\rangle=\frac{1}{4}\sum_{\substack{\lambda,\lambda^{\prime}\\s,s^{\prime}}}|\mathcal{M}_1|^2=\langle|\mathcal{M}_1^{(a)}|^2\rangle+\langle|\mathcal{M}_1^{(b)}|^2\rangle+\left\langle2\Re{\mathcal{M}_1^{(a)\dagger}\mathcal{M}_1^{(b)}}\right\rangle.\label{eq25}
\end{equation}
Choosing a time-like background vector $b^\mu=(b_0,0,0,0)$, we get
\begin{eqnarray}
        \langle|\mathcal{M}_1^{(a)}|^2\rangle&=&g^2e^2b_0^2F^2(\beta)\varepsilon_{\mu0\alpha\kappa} \varepsilon_{\omega0\phi\tau}g^{\tau\kappa}k^{\alpha}k^\phi\nonumber\\
        &&\times\Tr{(\slashed{p}^{\prime}+m)\gamma^\sigma S(p+k)\gamma^\mu(\slashed{p}+m)\gamma^\omega S^{\dagger}(p+k)\gamma_\sigma},\\
         \langle|\mathcal{M}_1^{(b)}|^2\rangle&=&g^2e^2b_0^2F^2(\beta)\varepsilon_{\mu0\alpha\kappa} \varepsilon_{\omega0\phi\tau}g^{\tau\kappa}(k^\prime)^{\alpha}(k^\prime)^\phi\nonumber\\
         &&\times\Tr{(\slashed{p}^{\prime}+m)\gamma^\sigma S(p-k^\prime)\gamma^\mu(\slashed{p}+m)\gamma^\omega S^{\dagger}(p-k^\prime)\gamma_\sigma}
\end{eqnarray}
and
\begin{eqnarray}
        \left\langle2\Re{\mathcal{M}_1^{(a)\dagger}\mathcal{M}_1^{(b)}}\right\rangle&=&2g^2e^2b_0^2F^2(\beta)\Re\{\varepsilon_{\mu0\alpha\kappa} \varepsilon_{\omega0\phi\tau}k^{\phi}(k^{\prime})^{\alpha}\nonumber\\
        &&\times\Tr[\slashed{p}\gamma^\omega S^\dagger(p+k)\gamma^\kappa\slashed{p}^{\prime}\gamma^\tau S(p-k^{\prime})\gamma^\mu]\}.
\end{eqnarray}
This corresponds to diagrams with one vertex that breaks the Lorentz symmetry.

In this way, the third term of Eq. (\ref{eq002}) becomes
\begin{equation}
    \langle \mathcal{M}_0\mathcal{M}_2^\dagger\rangle=\langle \mathcal{M}_0^{(a)}\mathcal{M}_2^{(a)\dagger}\rangle+\langle \mathcal{M}_0^{(a)}\mathcal{M}_2^{(b)\dagger}\rangle+\langle \mathcal{M}_0^{(b)}\mathcal{M}_2^{(a)\dagger}\rangle+\langle \mathcal{M}_0^{(b)}\mathcal{M}_2^{(b)\dagger}\rangle,\label{eq26}
\end{equation}
where
\begin{eqnarray}
       \langle \mathcal{M}_0^{(a)}\mathcal{M}_2^{(a)\dagger}\rangle &=&-\frac{g^2e^2b_0^2}{4}F^2(\beta)\varepsilon_{\mu0\alpha\kappa}\varepsilon_{\omega0\sigma\tau}k^{\sigma}(k^{\prime})^{\alpha}\nonumber\\
       &&\times\Tr[(\slashed{p}^{\prime}+m)\gamma^\kappa S(p+k)\gamma^\tau(\slashed{p}+m)\gamma^\omega S^{\dagger}(p+k)\gamma^\mu ],\\
        \langle \mathcal{M}_0^{(a)}\mathcal{M}_2^{(b)\dagger}\rangle&=&-\frac{g^2e^2b_0^2}{4}F^2(\beta)\varepsilon_{\mu0\alpha\kappa}\varepsilon_{\omega0\sigma\tau}(k^{\prime})^{\sigma}k^{\alpha}\nonumber\\
        &&\times\Tr[(\slashed{p}^{\prime}+m)\gamma^\tau S(p+k)\gamma^\kappa(\slashed{p}+m)\gamma^\omega S^{\dagger}(p-k^{\prime})\gamma^\mu],\\
         \langle \mathcal{M}_0^{(b)}\mathcal{M}_2^{(a)\dagger}\rangle&=&-\frac{g^2e^2b_0^2}{4}F^2(\beta)\varepsilon_{\mu0\alpha\kappa}\varepsilon_{\omega0\sigma\tau}k^{\sigma}(k^{\prime})^{\alpha}\nonumber\\
         &&\times\Tr[(\slashed{p}^{\prime}+m)\gamma^\tau S(p-k^{\prime})\gamma^\kappa(\slashed{p}+m)\gamma^\omega S^{\dagger}(p+k)\gamma^\mu],\\
          \langle \mathcal{M}_0^{(b)}\mathcal{M}_2^{(b)\dagger}\rangle&=&-\frac{g^2e^2b_0^2}{4}F^2(\beta)\varepsilon_{\mu0\alpha\kappa}\varepsilon_{\omega0\sigma\tau}(k^{\prime})^{\sigma}k^{\alpha}\nonumber\\
        &&\times\Tr[(\slashed{p}^{\prime}+m)\gamma^\kappa S(p-k^{\prime})\gamma^\tau(\slashed{p}+m)\gamma^\omega S^{\dagger}(p-k^{\prime})\gamma^\mu].
\end{eqnarray}
Note that the background vector $b^\mu$ is time-like as in the previous case. Although only the pure time-like case has been considered, it is interesting to observe that, if a pure space-like background is assumed, a stronger angular dependence in our results is expected, as shown in the references \cite{comptonviolation, bhabhaviolation}. 

In the next section, these results are combined to write the differential cross section for Compton scattering with Lorentz violation at finite temperature.

\section{Differential cross section}

Here the main objective is to calculate the differential cross section at finite temperature for the Compton scattering process with corrections due to the Lorentz violation. First, let us use the results found in the previous section to write the total probability density. Then using Eqs. (\ref{eq24}), (\ref{eq25}) and (\ref{eq26}) in Eq. (\ref{eq002}) and calculating the trace, the probability density is given as
\begin{equation}
        \langle |\mathcal{M}|^2\rangle=\left[e^4\left(\Gamma_1+\frac{2 \pi ^2}{\xi _a^2 \xi _b^2}\left(\Gamma_2+\Gamma_3\right)\right)+g^2e^2b_0^2\left(\Gamma_4+\frac{4\pi^2}{\xi _a^2 \xi _b^2}(\Gamma_5+\Gamma_6+\Gamma_7+\Gamma_8)\right)\right]F^2(\beta),
\end{equation}
where the functions $\Gamma_i$ with $i=1,...,8$ are defined, explicitly,  as
\begin{eqnarray}
 \Gamma_1&=&\frac{ \omega^{\prime}2 \cos{\theta}}{\omega }+\frac{2 \omega  \cos{\theta}}{\omega^{\prime}}+4 (\cos{\theta}-1)-\frac{16 \omega  \sin ^4\left(\frac{\theta
   }{2}\right) \omega^{\prime}}{m^2}-\frac{8  \left(\omega^{\prime}-\omega \right)\sin ^2\left(\frac{\theta }{2}\right)}{m}\nonumber\\
   &&+\frac{2 m \left(\omega -\omega^{\prime}\right)
   \left[\omega ^2+\left(\omega^{\prime}\right)^2\right]}{\omega ^2 \left(\omega^{\prime}\right)^2},
\end{eqnarray}
\begin{eqnarray}
    \Gamma_2&=& 2 \left(\frac{\delta \left(m+\omega -\xi _a\right)}{1+e^{\beta  (m+\omega )}}+\frac{\delta \left(m+\omega +\xi _a\right)}{1+e^{\beta 
   (m+\omega )}}\right)^2 \sin ^2\left(\frac{\theta }{2}\right) \xi _b^2 \omega^{\prime} \omega ^3-8 \left(\omega\omega^{\prime}\cos{\theta}\right) (\omega
   -\omega  \cos{\theta} )\nonumber\\
   &&\times \left(\frac{\delta \left(m+\omega -\xi _a\right)}{1+e^{\beta  (m+\omega )}}+\frac{\delta \left(m+\omega +\xi
   _a\right)}{1+e^{\beta  (m+\omega )}}\right) \left(\frac{\delta \left(m-\xi _b-\omega^{\prime}\right)}{1+e^{\beta  \left(m-\omega^{\prime}\right)}}+\frac{\delta
   \left(m+\xi _b-\omega^{\prime}\right)}{1+e^{\beta  \left(m-\omega^{\prime}\right)}}\right) \xi _a \xi _b \omega^{\prime}\nonumber\\
   &&+\xi _a^2 \Biggl\{(\omega -\omega  \cos\theta)
   \left(\frac{\delta \left(m-\xi _b-\omega^{\prime}\right)}{1+e^{\beta  \left(m-\omega^{\prime}\right)}}+\frac{\delta \left(m+\xi _b-\omega^{\prime}\right)}{1+e^{\beta 
   \left(m-\omega^{\prime}\right)}}\right)^2 \left(\omega^{\prime}\right)^3\nonumber\\
   &&-2 \omega  \Biggl[\frac{\delta^2 \left(m+\omega -\xi _a\right)}{\left(1+e^{\beta  (m+\omega
   )}\right)^2}+\frac{2 \left(-\frac{\delta \left(m+\omega +\xi _a\right)}{1+e^{\beta  (m+\omega )}}+\frac{4 \delta \left(m-\xi _b-\omega
   '\right)}{1+e^{\beta  \left(m-\omega^{\prime}\right)}}-\frac{4 \delta \left(m+\xi _b-\omega^{\prime}\right)}{1+e^{\beta  \left(m-\omega^{\prime}\right)}}\right) \delta
   \left(m+\omega -\xi _a\right)}{1+e^{\beta  (m+\omega )}}\nonumber\\
   &&+\frac{\delta^2 \left(m+\omega +\xi _a\right)}{\left(1+e^{\beta  (m+\omega
   )}\right)^2}+\frac{\delta^2 \left(m-\xi _b-\omega^{\prime}\right)}{\left(1+e^{\beta  \left(m-\omega^{\prime}\right)}\right)^2}+\frac{\delta^2 \left(m+\xi _b-\omega
   '\right)}{\left(1+e^{\beta  \left(m-\omega^{\prime}\right)}\right)^2}+\frac{8 \delta \left(m+\omega +\xi _a\right) \delta \left(m+\xi _b-\omega
   '\right)}{\left(1+e^{\beta  (m+\omega )}\right) \left(1+e^{\beta  \left(m-\omega^{\prime}\right)}\right)}\nonumber\\
   &&+\frac{2 \delta \left(m-\xi _b-\omega^{\prime}\right)
   \left(-\frac{4 \delta \left(m+\omega +\xi _a\right)}{1+e^{\beta  (m+\omega )}}-\frac{\delta \left(m+\xi _b-\omega^{\prime}\right)}{1+e^{\beta  \left(m-\omega
   '\right)}}\right)}{1+e^{\beta  \left(m-\omega^{\prime}\right)}}\Biggl] \sin ^2\left(\frac{\theta }{2}\right) \xi _b^2 \omega^{\prime}\Biggl\},\\
     \Gamma_3&=&2m \xi _a \xi _b
    \Biggl\{\xi _aE \Biggl[\frac{\delta^2 \left(m+\omega -\xi _a\right)}{\left(1+e^{\beta  (m+\omega )}\right)^2}-\frac{2 \delta \left(m+\omega +\xi
   _a\right) \delta \left(m+\omega -\xi _a\right)}{\left(1+e^{\beta  (m+\omega )}\right)^2}+\frac{\delta^2 \left(m+\omega +\xi
   _a\right)}{\left(1+e^{\beta  (m+\omega )}\right)^2}\nonumber\\
   &&+\frac{\delta^2 \left(m-\xi _b-\omega^{\prime}\right)}{\left(1+e^{\beta  \left(m-\omega
'\right)}\right)^2}+\frac{\delta^2 \left(m+\xi _b-\omega^{\prime}\right)}{\left(1+e^{\beta  \left(m-\omega^{\prime}\right)}\right)^2}-\frac{2 \delta \left(m-\xi
   _b-\omega^{\prime}\right) \delta \left(m+\xi _b-\omega^{\prime}\right)}{\left(1+e^{\beta  \left(m-\omega^{\prime}\right)}\right)^2} \xi _b\nonumber\\
   &&-\left(\frac{\delta
   \left(m-\xi _b-\omega^{\prime}\right)}{1+e^{\beta  \left(m-\omega^{\prime}\right)}}-\frac{\delta \left(m+\xi _b-\omega^{\prime}\right)}{1+e^{\beta  \left(m-\omega
   '\right)}}\right)\nonumber\\
   &&\times \left(\frac{\delta \left(m-\xi _b-\omega^{\prime}\right)}{1+e^{\beta  \left(m-\omega^{\prime}\right)}}+\frac{\delta \left(m+\xi _b-\omega
   '\right)}{1+e^{\beta  \left(m-\omega^{\prime}\right)}}\right) \omega^{\prime} \left(\omega^{\prime}-\omega  \cos{\theta}\right)\Biggl]\nonumber\\
   &&-\omega  \left(\frac{\delta
   \left(m+\omega -\xi _a\right)}{1+e^{\beta  (m+\omega )}}-\frac{\delta \left(m+\omega +\xi _a\right)}{1+e^{\beta  (m+\omega )}}\right)\nonumber\\
   &&\times \left(\frac{\delta
   \left(m+\omega -\xi _a\right)}{1+e^{\beta  (m+\omega )}}+\frac{\delta \left(m+\omega +\xi _a\right)}{1+e^{\beta  (m+\omega )}}\right) \xi _b
   \left(\omega -\omega^{\prime}\cos{\theta} \right)\Biggl\},\\
   \Gamma_4&=&2 \omega^{\prime} (-6 E \cos{\theta} +6 \omega  \cos{\theta}-3 \omega  \cos{2 \theta}+\omega )-2 [3 \omega  (\omega -2 E) \cos{\theta}]+\omega  (4
   E-\omega )\nonumber\\
   &&-\frac{8 \sin ^2\left(\frac{\theta }{2}\right) \left\{\omega^{\prime} [3 \omega  (E+\omega ) \cos{\theta}+\omega  (E-2 \omega )]-3 \omega  \cos{\theta} \left(\omega^{\prime}\right)^2+2 \omega ^3\right\}}{m}+\frac{8 \omega ^2 (E+\omega )}{\omega^{\prime}}\nonumber\\
   &&-2 (3 \cos{\theta}-1) \left(\omega^{\prime}\right)^2-\frac{48
   \omega ^2 \sin ^6\left(\frac{\theta }{2}\right) \left(\omega^{\prime}\right)^2}{m^2}-\frac{4 m \left(\omega -\omega^{\prime}\right) \left(3 \omega^{\prime}\cos{\theta} +2
   \omega \right)}{\omega^{\prime}},
\end{eqnarray}
\begin{eqnarray}  
     \Gamma_5&=&4 E m\omega ^2 \left[\frac{\delta \left(m+\omega -\xi _a\right)}{1+e^{\beta  (m+\omega
   )}}+\frac{\delta \left(m+\omega +\xi _a\right)}{1+e^{\beta  (m+\omega )}}\right]^2 \xi _b^2 \omega ^4\nonumber\\
   &&+6 (\cos{2\theta} )+3) \left[\frac{\delta \left(m+\omega -\xi _a\right)}{1+e^{\beta  (m+\omega )}}+\frac{\delta
   \left(m+\omega +\xi _a\right)}{1+e^{\beta  (m+\omega )}}\right]\nonumber\\
   &&\times\left[\frac{\delta \left(m-\xi _b-\omega
   '\right)}{1+e^{\beta  \left(m-\omega^{\prime}\right)}}+\frac{\delta \left(m+\xi _b-\omega^{\prime}\right)}{1+e^{\beta 
   \left(m-\omega^{\prime}\right)}}\right] \sin ^2\left(\frac{\theta }{2}\right) \xi _a \xi _b \left(\omega^{\prime}\right)^3
   \omega ^3\nonumber\\
   &&+16 \cos{\theta} \left[\frac{\delta \left(m+\omega -\xi _a\right)}{1+e^{\beta  (m+\omega )}}-\frac{\delta
   \left(m+\omega +\xi _a\right)}{1+e^{\beta  (m+\omega )}}\right]  \nonumber\\
   &&\times\left[\frac{\delta \left(m-\xi _b-\omega
   '\right)}{1+e^{\beta  \left(m-\omega^{\prime}\right)}}-\frac{\delta \left(m+\xi _b-\omega^{\prime}\right)}{1+e^{\beta 
   \left(m-\omega^{\prime}\right)}}\right] \sin ^2\left(\frac{\theta }{2}\right) \xi _a^2 \xi _b^2 \left(\omega
   '\right)^2 \omega ^2\nonumber\\
   &&-4 \cos{\theta} \xi _a^2 \left(\omega^{\prime}\right)^2 \Biggl\{2 \sin ^2\left(\frac{\theta
   }{2}\right) \left(\omega^{\prime}\right)^2 \left[\frac{\delta \left(m-\xi _b-\omega^{\prime}\right)}{1+e^{\beta 
   \left(m-\omega^{\prime}\right)}}+\frac{\delta \left(m+\xi _b-\omega^{\prime}\right)}{1+e^{\beta  \left(m-\omega
   '\right)}}\right]^2\nonumber\\
   &&+(\omega^{\prime})^{2} (\cos{\theta}-1) \left[\frac{\delta \left(m-\xi _b-\omega
   '\right)}{1+e^{\beta  \left(m-\omega^{\prime}\right)}}+\frac{\delta \left(m+\xi _b-\omega^{\prime}\right)}{1+e^{\beta 
   \left(m-\omega^{\prime}\right)}}\right]^2-2 \sin ^2\left(\frac{\theta }{2}\right) \xi _b^2\nonumber\\
   &&\times \left[\frac{\delta
   \left(m-\xi _b-\omega^{\prime}\right)}{1+e^{\beta  \left(m-\omega^{\prime}\right)}}-\frac{\delta \left(m+\xi _b-\omega
   '\right)}{1+e^{\beta  \left(m-\omega^{\prime}\right)}}\right] \left[\frac{\delta \left(m+\omega -\xi _a\right)}{1+e^{\beta  (m+\omega
   )}}-\frac{\delta \left(m+\omega +\xi _a\right)}{1+e^{\beta  (m+\omega )}}\right]\Biggl\},\\
     \Gamma_6&=&4 m\omega \xi _a^2 \Biggl\{\Biggl[E \omega  \Biggl(\frac{\delta^2 \left(m+\omega -\xi
   _a\right)}{\left(1+e^{\beta  (m+\omega )}\right)^2}-\frac{2 \delta \left(m+\omega +\xi _a\right) \delta
   \left(m+\omega -\xi _a\right)}{\left(1+e^{\beta  (m+\omega )}\right)^2}+\frac{\delta ^2\left(m+\omega +\xi
   _a\right)}{\left(1+e^{\beta  (m+\omega )}\right)^2}\nonumber\\
   &&+\frac{\delta^2 \left(m-\xi _b-\omega
   '\right)}{\left(1+e^{\beta  \left(m-\omega^{\prime}\right)}\right)^2}+\frac{\delta^2 \left(m+\xi _b-\omega
   '\right)}{\left(1+e^{\beta  \left(m-\omega^{\prime}\right)}\right)^2}-\frac{2 \delta \left(m-\xi _b-\omega
   '\right) \delta \left(m+\xi _b-\omega^{\prime}\right)}{\left(1+e^{\beta  \left(m-\omega^{\prime}\right)}\right)^2}\Biggl)\nonumber\\
   &&-2
   E \cos{\theta} \left(\frac{\delta \left(m+\omega -\xi _a\right)}{1+e^{\beta  (m+\omega )}}-\frac{\delta
   \left(m+\omega +\xi _a\right)}{1+e^{\beta  (m+\omega )}}\right) \left(\frac{\delta \left(m-\xi _b-\omega
   '\right)}{1+e^{\beta  \left(m-\omega^{\prime}\right)}}-\frac{\delta \left(m+\xi _b-\omega^{\prime}\right)}{1+e^{\beta 
   \left(m-\omega^{\prime}\right)}}\right) \omega^{\prime}\Biggl] \xi _b^2\nonumber\\
   &&+\left(\frac{\delta \left(m-\xi _b-\omega
   '\right)}{1+e^{\beta  \left(m-\omega^{\prime}\right)}}+\frac{\delta \left(m+\xi _b-\omega^{\prime}\right)}{1+e^{\beta 
   \left(m-\omega^{\prime}\right)}}\right) \omega^{\prime} \Biggl[2 \omega ^2 \cos{\theta} \left(\frac{\delta \left(m-\xi
   _b-\omega^{\prime}\right)}{1+e^{\beta  \left(m-\omega^{\prime}\right)}}-\frac{\delta \left(m+\xi _b-\omega
   '\right)}{1+e^{\beta  \left(m-\omega^{\prime}\right)}}\right)\nonumber\\
   &&-\Biggl(-\frac{E \delta \left(m+\omega -\xi
   _a\right)}{1+e^{\beta  (m+\omega )}}+\frac{E \delta \left(m+\omega +\xi _a\right)}{1+e^{\beta  (m+\omega )}}+E
   \cos{\theta} \left(\frac{\delta \left(m+\omega -\xi _a\right)}{1+e^{\beta  (m+\omega )}}-\frac{\delta
   \left(m+\omega +\xi _a\right)}{1+e^{\beta  (m+\omega )}}\right)\nonumber\\
   &&+\frac{2 \omega  \delta \left(m-\xi _b-\omega
   '\right)}{1+e^{\beta  \left(m-\omega^{\prime}\right)}}-\frac{2 \omega  \delta \left(m+\xi _b-\omega
   '\right)}{1+e^{\beta  \left(m-\omega^{\prime}\right)}}\Biggl) \omega^{\prime}\Biggl] \xi _b\nonumber\\
   &&+E \omega  \left(\frac{\delta
   \left(m-\xi _b-\omega^{\prime}\right)}{1+e^{\beta  \left(m-\omega^{\prime}\right)}}+\frac{\delta \left(m+\xi _b-\omega
   '\right)}{1+e^{\beta  \left(m-\omega^{\prime}\right)}}\right)^2 \left(\omega^{\prime}\right)^2\Biggl\}
\end{eqnarray}
\begin{eqnarray}
    \Gamma_7&=&2 E m \omega^{\prime}\omega \Biggl\{\Biggl[\cos{\theta} \Biggl(\frac{\delta \left(m+\omega -\xi _a\right)}{1+e^{\beta  (m+\omega
   )}}-\frac{\delta \left(m+\omega +\xi _a\right)}{1+e^{\beta  (m+\omega )}}+\frac{\delta \left(m-\xi _b-\omega
   '\right)}{1+e^{\beta  \left(m-\omega^{\prime}\right)}}\nonumber\\
   &&-\frac{\delta \left(m+\xi _b-\omega^{\prime}\right)}{1+e^{\beta 
   \left(m-\omega^{\prime}\right)}}\Biggl)^2 \xi _b^2
 +\Biggl(-\frac{\delta \left(m+\omega -\xi _a\right)}{1+e^{\beta 
   (m+\omega )}}+\frac{\delta \left(m+\omega +\xi _a\right)}{1+e^{\beta  (m+\omega )}}\nonumber\\
   &&+\cos{\theta}
   \left(\frac{\delta \left(m+\omega -\xi _a\right)}{1+e^{\beta  (m+\omega )}}-\frac{\delta \left(m+\omega +\xi
   _a\right)}{1+e^{\beta  (m+\omega )}}\right)
  +\frac{2 \delta \left(m-\xi _b-\omega^{\prime}\right)}{1+e^{\beta 
   \left(m-\omega^{\prime}\right)}}\nonumber\\
   &&-\frac{2 \delta \left(m+\xi _b-\omega^{\prime}\right)}{1+e^{\beta  \left(m-\omega
   '\right)}}\Biggl) \left(\frac{\delta \left(m-\xi _b-\omega^{\prime}\right)}{1+e^{\beta  \left(m-\omega
   '\right)}}+\frac{\delta \left(m+\xi _b-\omega^{\prime}\right)}{1+e^{\beta  \left(m-\omega^{\prime}\right)}}\right) \omega^{\prime}
   \xi _b\nonumber\\
   &&+\cos{\theta} \left(\frac{\delta \left(m-\xi _b-\omega^{\prime}\right)}{1+e^{\beta  \left(m-\omega
   '\right)}}+\frac{\delta \left(m+\xi _b-\omega^{\prime}\right)}{1+e^{\beta  \left(m-\omega^{\prime}\right)}}\right)^2
   \left(\omega^{\prime}\right)^2\biggl] \xi _a^2\nonumber\\
   &&+\omega  \left(\frac{\delta \left(m+\omega -\xi _a\right)}{1+e^{\beta 
   (m+\omega )}}+\frac{\delta \left(m+\omega +\xi _a\right)}{1+e^{\beta  (m+\omega )}}\right) \xi _b\nonumber\\
   &&\times \Biggl[2
   \left(\frac{\delta \left(m-\xi _b-\omega^{\prime}\right)}{1+e^{\beta  \left(m-\omega^{\prime}\right)}}+\frac{\delta
   \left(m+\xi _b-\omega^{\prime}\right)}{1+e^{\beta  \left(m-\omega^{\prime}\right)}}\right) \omega^{\prime} \cos ^2(\theta
   )+\nonumber\\
   &&\Biggl(-\frac{2 \delta \left(m+\omega -\xi _a\right)}{1+e^{\beta  (m+\omega )}}+\frac{2 \delta \left(m+\omega
   +\xi _a\right)}{1+e^{\beta  (m+\omega )}}+\frac{\delta \left(m-\xi _b-\omega^{\prime}\right)}{1+e^{\beta 
   \left(m-\omega^{\prime}\right)}}-\frac{\delta \left(m+\xi _b-\omega^{\prime}\right)}{1+e^{\beta  \left(m-\omega
   '\right)}}\nonumber\\
   &&-\cos{\theta} \left(\frac{\delta \left(m-\xi _b-\omega^{\prime}\right)}{1+e^{\beta  \left(m-\omega
   '\right)}}-\frac{\delta \left(m+\xi _b-\omega^{\prime}\right)}{1+e^{\beta  \left(m-\omega^{\prime}\right)}}\right)\Biggl)
   \xi _b\Biggl] \xi _a\nonumber\\
   &&+\omega ^2 \cos (\theta ) \left(\frac{\delta \left(m+\omega -\xi _a\right)}{1+e^{\beta 
   (m+\omega )}}+\frac{\delta \left(m+\omega +\xi _a\right)}{1+e^{\beta  (m+\omega )}}\right)^2 \xi _b^2\Biggl\},\\   
    \Gamma_8&=&m \left[\frac{\delta \left(m+\omega -\xi _a\right)}{1+e^{\beta  (m+\omega )}}+\frac{\delta
   \left(m+\omega +\xi _a\right)}{1+e^{\beta  (m+\omega )}}\right] \xi _a \xi _b\nonumber\\
   &&\times \Biggl\{-8 E \omega ^2 \cos
   ^2{\theta} \left(\frac{\delta \left(m-\xi _b-\omega^{\prime}\right)}{1+e^{\beta  \left(m-\omega
   '\right)}}+\frac{\delta \left(m+\xi _b-\omega^{\prime}\right)}{1+e^{\beta  \left(m-\omega^{\prime}\right)}}\right)
   \left(\omega^{\prime}\right)^2\nonumber\\
   &&-4 \omega  \xi _b \Biggl[2 \omega ^3 \left(\frac{\delta \left(m+\omega -\xi
   _a\right)}{1+e^{\beta  (m+\omega )}}-\frac{\delta \left(m+\omega +\xi _a\right)}{1+e^{\beta  (m+\omega
   )}}\right)\nonumber\\
   &&-\omega  \Biggl(\cos{\theta} \Biggl(2 \omega  \left(\frac{\delta \left(m+\omega -\xi
   _a\right)}{1+e^{\beta  (m+\omega )}}-\frac{\delta \left(m+\omega +\xi _a\right)}{1+e^{\beta  (m+\omega
   )}}\right)\nonumber\\
   &&+\frac{E \delta \left(m-\xi _b-\omega^{\prime}\right)}{1+e^{\beta  \left(m-\omega^{\prime}\right)}}-\frac{E \delta
   \left(m+\xi _b-\omega^{\prime}\right)}{1+e^{\beta  \left(m-\omega^{\prime}\right)}}\Biggl)\nonumber\\
   &&-E \left(\frac{\delta \left(m-\xi
   _b-\omega^{\prime}\right)}{1+e^{\beta  \left(m-\omega^{\prime}\right)}}-\frac{\delta \left(m+\xi _b-\omega
   '\right)}{1+e^{\beta  \left(m-\omega^{\prime}\right)}}\right)\Biggl) \omega^{\prime}\Biggl]\Biggl\}.
\end{eqnarray}
In these expressions, only the first-order terms of the electron mass are considered, since it is a very small quantity. Note that in most terms there are products of delta functions, and in general these products have the same argument, this is due to the formalism being used \cite{delta1,delta2}. However, to avoid divergences, there are ways to  deal with this situation, i.e., one can work with regularized forms of these functions, as shown in reference \cite{realandimaginary}.

Using the definition of the differential cross section in the laboratory frame  \cite{peskin}, that is,
\begin{equation}
\frac{d\sigma}{d\Omega}=\frac{(\omega^\prime)^2}{16\pi^2(s-m^2)^2}\langle |\mathcal{M}|^2\rangle,
\end{equation}
for Compton scattering violating Lorentz symmetry at finite temperature, the following result is obtained
\begin{eqnarray}
        \left(\frac{d\sigma_{LV}}{d\Omega}\right)_\beta&=&\frac{e^4(\omega^\prime)^2}{16\pi^2(s-m^2)^2}\left[\Gamma_1+\frac{2 \pi ^2}{\xi _a^2 \xi _b^2}\left(\Gamma_2+\Gamma_3\right)\right]F^2(\beta)\nonumber\\
        &&+\frac{g^2e^2b_0^2(\omega^\prime)^2}{16\pi^2(s-m^2)^2}\left[\Gamma_4+\frac{4\pi^2}{\xi _a^2 \xi _b^2}(\Gamma_5+\Gamma_6+\Gamma_7+\Gamma_8)\right]F^2(\beta).
\end{eqnarray}
It is interesting to note that there are corrections due to both the Lorentz violation and the finite temperature.  Let us analyze this result for some limits. At the high temperatures limit, i.e. $T\to\infty$, leads to $\beta\to0$ implying $e^{\beta(m+\omega)}\to1$ and $e^{\beta(m-\omega^{\prime})}\to1$. In this way, the contribution $F^2(\beta)$ becomes the main term, then the temperature effects are dominant and very expressive. On the other hand, at the low temperature limit $T\to0$ we get $\beta\to\infty$ and therefore
\begin{equation}
   \lim_{\beta\to\infty}\frac{1}{\left[e^{\beta  (m+\omega )}+1\right]}=0;\quad\quad\text{and}\quad\quad\lim_{\beta\to\infty}\frac{1}{\left[e^{\beta  \left(m-\omega^{\prime}\right)}+1\right]}=0.
\end{equation}
Furthermore, except for $i=1,4$, the functions $\Gamma_i$ go to zero quickly, while $F^2(\beta)=1$. Then the differential cross-section becomes
\begin{equation}
    \begin{split}
        \frac{d\sigma_{LV}}{d\Omega}&=\frac{e^4(\omega^\prime)^2}{16\pi^2(s-m^2)^2}\Gamma_1+\frac{g^2e^2b_0^2(\omega^\prime)^2}{16\pi^2(s-m^2)^2}\Gamma_4.\label{eq23}
    \end{split}
\end{equation}
This is the differential cross-section for Compton scattering with Lorentz violation at zero temperature. In addition, taking $gb_0=0$ in Eq. (\ref{eq23}) leads to the usual result for Compton scattering, which, expanding $\Gamma_1$, can be written as the Klein-Nishina formula given as
\begin{equation}
   \frac{d\sigma}{d(\cos{\theta})}=\frac{\pi\alpha^2}{m^2}\left(\frac{\omega^\prime}{\omega}\right)^2\left[\frac{\omega^\prime}{\omega}+\frac{\omega}{\omega^\prime}-\sin^2{\theta}\right],\label{eq27}
\end{equation}
with $\alpha$ being the fine-structure constant. The solid angle element $d\Omega=2\pi\sin{\theta}d\theta$ and the relation
\begin{equation}
    \omega^{\prime}=\frac{\omega}{1+\frac{\omega}{m}(1-\cos{\theta})}
\end{equation}
have been used \cite{klein}.

\begin{figure}[ht]
    \centering
    \includegraphics[scale=0.8]{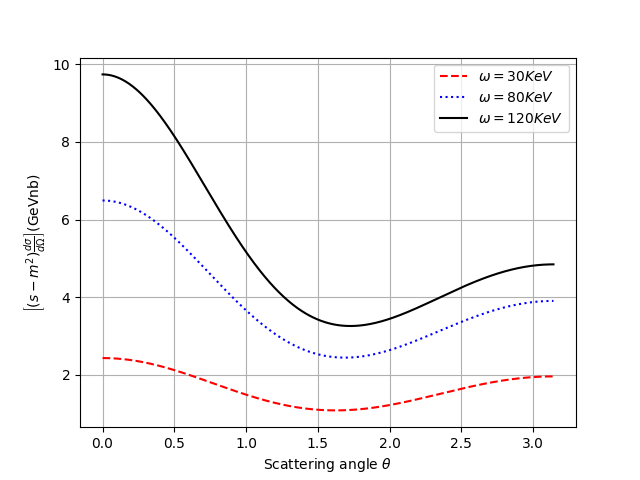}
    \caption{The Lorentz-invariant part of the differential cross section multiplied by $(s-m^2)$ at zero temperature, in terms of the scattering angle, for $\omega=30KeV$ (red dashed line), $\omega=80KeV$ (blue dotted line) and $\omega=120KeV$ (black solid line).}
    \label{fig3}
    \end{figure}

From Eq. (\ref{eq23}),  the part of the differential cross section without Lorentz violation at zero temperature, that is, Eq. (\ref{eq27}) is plotted in Figure \ref{fig3}. It can be seen that this quantity strongly depends on the initial frequency of the photon. Furthermore, the minimum of this function is around $\pi/2$, so this situation, from Figure \ref{fig2}, means that the least likely event is one in which the scattered photon moves in a normal direction with respect to the incident. For this differential cross section, the normalization factor $(s-m^2)$ in units of $GeV$  is used, and the nanobar is also used for the units of area. 

\begin{figure}[ht]
\centering
    \includegraphics[scale=0.8]{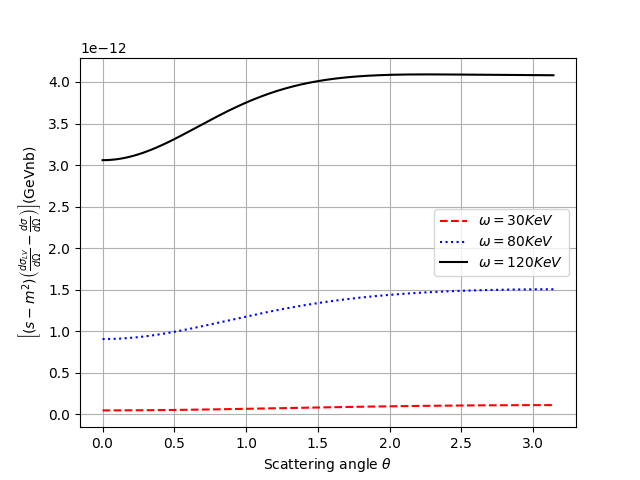}
     \caption{The violating-Lorentz part of the differential cross section multiplied by $(s-m^2)$ at zero temperature, in terms of the scattering angle, for $\omega=30KeV$ (red dashed line), $\omega=80KeV$ (blue dotted line) and $\omega=120KeV$ (black solid line).}
    \label{fig4}
    \end{figure}

On the other hand, the contribution of symmetry breaking to the total cross section at zero temperature, of Eq. (\ref{eq23}) and Eq. (\ref{eq27}), is given by
\begin{equation}
    \frac{d\sigma_{LV}}{d\Omega}-\frac{d\sigma}{d\Omega}=\frac{g^2e^2b_0^2(\omega^\prime)^2}{16\pi^2(s-m^2)^2}\Gamma_4.\label{eq28}
\end{equation}
This function is shown in Figure \ref{fig4}. Like Eq. (90), Eq. (92) also depends on both the incident photon energy and the scattering angle. However, the energy dependence in the second case is more expressive, so that for high energies this implies a greater correction in the total cross section due to the Lorentz violation. It can also be noted that this scale is on the order of $10^{-12}$, and this fact shows that this correction is very small and makes the measurement processes need to be more and more accurate.

The Figures \ref{fig4} and \ref{fig3} show that there is an interesting point around $\pi/2$. In the first case, this point was the minimum spot, now it divides the regime where the function starts to have an apparently constant behavior. For these pictures, the correct units were recovered by inserting the factor $(\hbar c)^2$. In addition, $gb_0=10^{-3}(GeV)^{-1}$ is used  as proposed in \cite{bhabhaviolation,comptonviolation}.

\section{Conclusion}

Compton scattering with Lorentz violation at finite temperature has been analyzed. Temperature effects have been introduced by the TFD formalism, which is special because it gives us many calculations in a very similar way to those of quantum field theory at zero temperature. And although the solution obtained is very large and relatively complicated, it can give us a reasonable approximation of a real situation. The asymptotic bounds show that the relevant thermal effects emerge when the factor $\beta$ becomes null. This is because the Bose distribution turns out to be divergent. This photon condensate makes the differential cross section, at high temperatures, many times greater than at zero temperature.  Furthermore, the Lorentz violation also modifies the quantities obtained for this scattering. This symmetry breaking appears in the theory through a constant background field introduced by the covariant derivative. And this change is strongly dependent on the scattering angle and the initial energy of the photon, as shown in Figure \ref{fig4}. In other words, the smaller the $gb_0$ factor, the greater the initial energy must be. Therefore, this effect would be clearly observed when the energies involved are on the order of the Planck scale, i.e. as at the beginning of the universe for example, where quantum field theory and general relativity are expected to be unified.  In addition, it is important to emphasize that the study developed here differs subtly from that one analyzed in \cite{comptonviolation} at zero temperature. This difference arises because the non-minimal coupling term that leads to the Lorentz violation is different. Therefore, a direct comparison between the two results is not possible, although the dependence on Lorentz violation is similar, i.e. the correction due to the Lorentz violation is proportional to $b_0^2$.

\section*{Acknowledgments}

This work by A. F. S. is partially supported by National Council for Scientific and Technological Develo\-pment - CNPq project No. 313400/2020-2.

\section*{Data Availability Statement}

No Data associated in the manuscript.


\global\long\def\link#1#2{\href{http://eudml.org/#1}{#2}}
 \global\long\def\doi#1#2{\href{http://dx.doi.org/#1}{#2}}
 \global\long\def\arXiv#1#2{\href{http://arxiv.org/abs/#1}{arXiv:#1 [#2]}}
 \global\long\def\arXivOld#1{\href{http://arxiv.org/abs/#1}{arXiv:#1}}


\end{document}